\documentclass[manuscript]{aastex}
\usepackage{txfonts}
\usepackage{latexsym,bm}
\usepackage{lineno}
\usepackage{multirow}
\slugcomment{Not to appear in Nonlearned J., 45.}

\shorttitle{Determining the onset time of solar energetic particles}

\shortauthors{Wang and Qin }

\begin{document}


\title{ESTIMATION OF THE RELEASE TIME OF SOLAR ENERGETIC PARTICLES NEAR THE SUN}

\author{ Yang Wang\altaffilmark{1} and Gang Qin\altaffilmark{1}}

\email{ywang@spaceweather.ac.cn; gqin@spaceweather.ac.cn}

\altaffiltext{1}{State Key Laboratory of Space Weather,  Center for Space Science and Applied Research, Chinese Academy of Sciences, Beijing 100190, China}

\begin{abstract}

This paper investigates the onset time of Solar Energetic Particle (SEP)  events with numerical simulations, and analyses the accuracy of the Velocity Dispersion Analysis (VDA) method. Using a 3-dimensional focused transport model, we calculate the fluxes of protons observed {\bf{ in the ecliptic}} at $1$ AU in {\bf{the}} energy range between $10$ MeV and $80$ MeV. 
In particular, three models are used to describe different SEP sources produced by flare or coronal shock, and the effects of particle perpendicular diffusion in the interplanetary space are also studied.
We have the following findings: When the observer is  disconnected from the source, the effects of perpendicular diffusion in the interplanetary space and  particles propagating in the solar atmosphere have significant influence on the VDA results. As a result, although the VDA method is valid with {\bf{impulsive}} source duration, low background, {\bf{and}}  weak scattering in the interplanetary space or fast diffusion in the solar atmosphere, the method is not valid with {\bf{gradual}} source duration, high background, or strong scattering.


\end{abstract}


\section{INTRODUCTION}

Solar Energetic Particles (SEPs) were first reported by \cite{forbush1946three}. Since coronal mass ejection (CME) was not discovered at that time,
SEPs were assumed to be accelerated by solar flare.
If this holds true, it is reasonable to assume that the size of SEPs source is close to that of flare. However, some SEP events could be simultaneously observed by multi-spacecraft with very wide spatial distribution which could be much wider than the size of flare. In order to interpret this phenomenon, two scenarios were proposed: (1)
particles can cross {\bf{ magnetic field }} lines in the interplanetary space with perpendicular diffusion \citep{mckibben1972azimuthal, Dresing2012SoPh,Dresing2014A&A}; and (2) particles can propagate in the solar atmosphere \citep{wibberenz1989coronal,Dresing2014A&A}. But later the SEP community realized that CMEs are important for particle acceleration, especially in  large SEP events  \citep{mason1984temporal,Gosling1993The,Zank2000JGR...10525079Z,li2003energetic}.
As a result, besides the former two scenarios, the third one was proposed: {\bf{The}} wide spread of SEPs can be explained by that SEPs are accelerated by large scale shocks. However, large SEP events are usually associated with {\bf{both}} flares and CMEs, so the role of flare and shock in {\bf{the}} acceleration process of SEPs is still in debate.  For historical development of the studies on SEP source, please refer to the review articles by \cite{Reames1999SSRv90413R} and \cite{reames2013two}.

Because of the effects of particle transport, it is difficult to distinguish {\bf{the}} signatures of different accelerators in
SEP fluxes at 1 AU. However, SEP fluxes observed in the interplanetary space show velocity
dispersion at the onset time. The Velocity Dispersion Analysis (VDA) method
has been used widely to investigate the SEP acceleration and
transport {\bf{processes}}
\citep{Krucker1999ApJ519864K,Krucker2000ApJ542L61K,kahler2006ApJ,Reames2009ApJ706844R,Tan2012ApJ,
li2013electron,Ding2014ApJ}.
This method assumes that the first arriving particles move along the
magnetic field lines, and the path length {\bf{travelled}} by particles between the source and observer is independent on energy. With these
assumptions, the SEP release time near the Sun and the
interplanetary path length can be determined by using the onset {\bf{times}} of different energy particles. {\bf{In addition}}, to compare the SEP release time with the electromagnetic signature of SEP source, the SEP source can be identified .

It has been known for a long time that when the observer is
far from the magnetic connection point of the source on the Sun,  the onset time of SEPs' flux shows a delay \citep{van1975variation, MaSung1978Interplanetary}.
This would  lead to changes in the results of VDA method.
\cite{Krucker1999ApJ519864K} calculated the release time near the Sun and path length in the interplanetary space of SEPs with $12$ short electron events observed by {\bf{the}}  Wind
spacecraft. The result of VDA method indicates two kinds of electron events. In the first kind of events, the electron release time is extremely close to the onset of a radio type III burst when the observer is connected to the flare. In the second kind of events, the electrons are released much later (e.g., half an hour ) than the onset of the type III burst when the observer is disconnected from the flare.  \cite{Huttunen2005A&A442673H} studied the release time of MeV/n protons and heliums observed by SOHO/ERNE. They found that the delay in SEPs release time derived from VDA method is related to the poor magnetic connection between the flare site and the spacecraft. For extremely high energy particle events, \cite{Reames2009ApJ706844R} studied the onset time of ion fluxes in ground-level enhancements. They {\bf{concluded that }} the time difference between the solar particle
release time and the onset time of metric type II radio burst  increases with the  angular distance of the observer's magnetic foot-point and the source increase.

According to different heliographic latitude observation,
\cite{Zhang2003JGRA1081154Z} analyzed an SEP event simultaneously observed by {\bf{the}} Ulysses and GOES spacecraft. The GOES spacecraft is located {\bf{ in the ecliptic}}, while the Ulysses is located at
$62^\circ$ South. The {\bf{SEP}} release time derived from GOES data is consistent with the onset of soft X-ray flux, and the path length is also
close to the Parker spiral. In contrast, the release time
derived from Ulysses data is $3$ hours later than the onset of soft
X-ray, and the path length is much longer than  Parker
spiral. {\bf{Further studies}} were  done by
\cite{dalla2003AnGeo21.1367D,Dalla2003GeoRL30sULY9D}, who analysed
$9$ SEP events observed by Ulysses at high latitudes
among which  $8$ events are observed at {\bf{latitudes}} more than $60^\circ$, and {\bf{the rest}} event is observed at $47.9^\circ$ latitude.
\cite{dalla2003AnGeo21.1367D,Dalla2003GeoRL30sULY9D} found that the path lengths {\bf{derived from the Ulysses data}} are
$1.06$ to $2.45$ times the length of Parker spiral, and the particle
release {\bf{ times are}} between $100$ and $350$ min later than {\bf{that}}
derived from {\bf{the}} SOHO and Wind measurements. The delay in particle
release time increases with the latitudinal difference  $\Delta \theta$ between
the spacecraft and the flare. 
{\bf{Based on the delay of particle release time derived from in-ecliptic measurements relative to that from high latitudes measurements}}, we can
conclude that {\bf{such kind of delay}} is related to the poor
connection between the source and spacecraft.

In order to use the VDA method more reasonably, many studies have been done to investigate the validity of {\bf {the}} method. The following are
their main conclusions: Firstly, when the parallel mean free path (MFP) is large
enough ($\lambda_{\parallel}>0.3 $ AU), interplanetary scattering has
only a small effect on the derived solar release time
\citep{Kallenrode1990ICRC5229K,Lintunen2004A&A420343L,diaz2011delay}.
Secondly, when the background level is below $0.01\%$ of the peak
intensity of flux, the onset time of SEP event can be determined
quite accurately \citep{Saiz2005ApJ...626.1131S}. In the above
works, the interplanetary scattering and background level effects in
the release time have been studied in detail. 
In this paper, we study how different source models and perpendicular diffusion {\bf{affect}} the onset time of SEP and the VDA results.  In section 2 we describe the SEP transport model. In Section 3 we show the simulation results. In Section 4 we discuss  {\bf{and summarize our results.}}

\section{MODEL}

We model the transport of SEPs following previous research
\citep[e.g.,][]{Qin2006JGRA..11108101Q,Zhang2009ApJ...692..109Z,droge2010ApJ,he2011propagation, zuo2011,wang2012effects,
qin2013transport,zuo2013acceleration,Wang2014ApJ789157W}.
A three-dimensional focused transport equation is written as
\citep{Skilling1971ApJ...170..265S,schlickeiser2002cosmic,
Qin2006JGRA..11108101Q,Zhang2009ApJ...692..109Z}
\begin{eqnarray}
  \frac{{\partial f}}{{\partial t}} = \nabla\cdot\left( \bm
  {\kappa_\bot}
\cdot\nabla f\right)- \left(v\mu \bm{\mathop b\limits^ \wedge}
+ \bm{V}^{sw}\right)
\cdot \nabla f + \frac{\partial }{{\partial \mu }}\left(D_
{\mu \mu }
\frac{{\partial f}}{{\partial \mu }}\right) \nonumber \\
  + p\left[ {\frac{{1 - \mu ^2 }}{2}\left( {\nabla  \cdot \bm{V}^
  {sw}  -
\bm{\mathop b\limits^ \wedge  \mathop b\limits^ \wedge } :\nabla
\bm{V}^{sw} } \right) +
\mu ^2 \bm{\mathop b\limits^ \wedge  \mathop b\limits^ \wedge}  :
\nabla \bm{V}^{sw} }
\right]\frac{{\partial f}}{{\partial p}} \nonumber \\
  - \frac{{1 - \mu ^2 }}{2}\left[ { - \frac{v}{L} + \mu \left
  ( {\nabla  \cdot
\bm{V}^{sw}  - 3\bm{\mathop b\limits^ \wedge  \mathop b\limits^
\wedge}  :\nabla \bm{V}^{sw} }
\right)} \right]\frac{{\partial f}}{{\partial \mu }},\label{dfdt}
\end{eqnarray}
where $f(\bm{x},\mu,p,t)$ is the gyrophase-averaged distribution
function, $\bm{x}$ is the position in a non-rotating heliographic
coordinate system; $p$, $\mu$,  and $v$ are the
momentum,  particle pitch-angle
cosine, and speed, respectively, in the solar wind frame; $t$ is
the time;
$\bm{V}^{sw}=V^{sw}\bm{\mathop r\limits^ \wedge}$ is the solar
wind velocity; $\bm{\mathop b\limits^ \wedge}$ is a unit vector
along the
local magnetic field, and $L$ is the magnetic focusing length given
by
$L=\left(\bm{\mathop b\limits^ \wedge}\cdot\nabla\\{ln} B_0
\right)^{-1}$ with $B_0$ being the magnitude of the background  magnetic field.
This equation includes many important particle transport effects such as
particle streaming along the field line, magnetic
focusing in the diverging IMF, adiabatic cooling in the
expanding solar wind, and the diffusion coefficients parallel and
perpendicular to the IMF.  Here, we use the Parker field model for the IMF, and the solar wind speed is  $400$ km/s.

The relationship of the $D_{\mu \mu }$ and parallel mean free path
$\lambda _\parallel$ is written as
\citep{Jokipii1966ApJ...146..480J,hasselmann1968scattering,earl1974diffusive}
\begin{equation}
\lambda _\parallel   = \frac{{3\upsilon}}{8}\int_{ - 1}^{ + 1}
{\frac{{(1 - \mu ^2 )^2 }}{{D_{\mu \mu } }}d\mu
},\label{lambda_parallel_1}
\end{equation}
and parallel diffusion coefficient $\kappa_\parallel$ can be written as $\kappa_\parallel=v\lambda_\parallel/3$.

We follow the model of pitch angle
diffusion coefficient from \citet{Beeck1986ApJ...311..437B},
see also \citep{qin2005model}
\begin{equation}
 D_{\mu \mu }  = D_0 \upsilon p^{ q-2} \left\{ {\left. {\left| \mu  \right.} \right|^{q - 1}  + h}
\right\}\left( {1 - \mu ^2 } \right),\label{D_mu_mu}
\end{equation}
where the constant $D_0$ controls the magnetic field fluctuations
level. The constant $q$ is chosen $5/3$ for a Kolmogorov spectrum
type of the power spectral density of magnetic field turbulence in
the inertial range. Furthermore, $h=0.01$ is chosen for non-linear effect of pitch angle diffusion at $\mu=0$ in the solar wind \citep{qin2009pitch,qin2014detailed}.

The relation of the particle momentum  and the
perpendicular diffusion coefficient is set as
\begin{equation}
{{\bm{\kappa }}_ \bot } = {\kappa _0}{\left( {\frac{p}{{1{\kern 2pt}  \textrm{GeV}{\kern 2pt}  { c^{ - 1} }}}} \right)^{1/3}}\left( {{\bf{I}} - \mathop {\bf{b}}\limits^ \wedge  \mathop {\bf{b}}\limits^ \wedge  } \right)\label{kappa_per}
\end{equation}
where $p$ is  particle momentum. { \bf{Different perpendicular 
diffusion coefficients could be obtained by altering $\kappa _0$, and  $ \kappa _ \bot/ \kappa _\parallel$ is set to $0.01$ { \bf{ in the ecliptic  at 1 AU}}  in our simulations.}} Note that we use
this formula for {\bf{the purpose of simplicity}}.  There are {\bf{some}} more complete models which are developed to describe the {\bf{particle}} diffusion in magnetic turbulence, for example, the  Non-Linear Guiding Center Theory (NLGC) \citep{Matthaeus2003ApJ...590L..53M,qin2014modification}.

We use boundary values to model the particle injection from the source, {\bf{which}} is chosen as the following form
\begin{equation}
{f_b}(z \le 0.05{\rm{AU}},{E_k},\theta ,\varphi ,t) = a\frac{{E_k^{ - \gamma }}}{{{p^2}}}\xi \left( {t,\theta ,\varphi } \right),\label{boundaryCondition}
\end{equation}
\[\xi \left( {t,\theta ,\varphi } \right) = \left\{ {\begin{array}{*{20}{c}}
{\frac{1}{t}\exp \left[ { - \frac{{{t_c}}}{t}{{\left( {\frac{{{\phi _s}}}{{{\phi _0}}}} \right)}^2} - \frac{t}{{{t_l}}}} \right]H\left( {{\phi _s} - \left| {\phi \left( {\theta ,\varphi } \right)} \right|} \right){\kern 50pt}{\textrm{Case 1}}},\\
{\begin{array}{*{20}{l}}
{\left\{ {\frac{1}{t}\exp \left[ { - \frac{{{t_c}}}{t}{{\left( {\frac{{{\phi _s}}}{{{\phi _0}}}} \right)}^2} - \frac{t}{{{t_l}}}} \right]H\left( {{\phi _s} - \left| {\phi \left( {\theta ,\varphi } \right)} \right|} \right) + } \right.}\\
{\left. {\frac{1}{t}\exp \left[ { - \frac{{{t_c}}}{t}{{\left( {\frac{{\phi \left( {\theta ,\varphi } \right)}}{{{\phi _0}}}} \right)}^2} - \frac{t}{{{t_l}}}} \right]\left[ {1 - H\left( {{\phi _s} - \left| {\phi \left( {\theta ,\varphi } \right)} \right|} \right)} \right]} \right\}{\kern 10pt}{\textrm{Case 2}}},
\end{array}}\\
{\frac{1}{t}\exp \left( { - \frac{{{t_c}}}{t} - \frac{t}{{{t_l}}}} \right)\exp \left( { - \frac{{\left| {\phi \left( {\theta ,\varphi } \right)} \right|}}{{{\phi _0}}}} \right)H\left( {{\phi _s} - \left| {\phi \left( {\theta ,\varphi } \right)} \right|} \right){\kern 20pt}{\textrm{Case 3}}}.
\end{array}} \right.\]

\begin{figure}
\epsscale{1.} \plotone{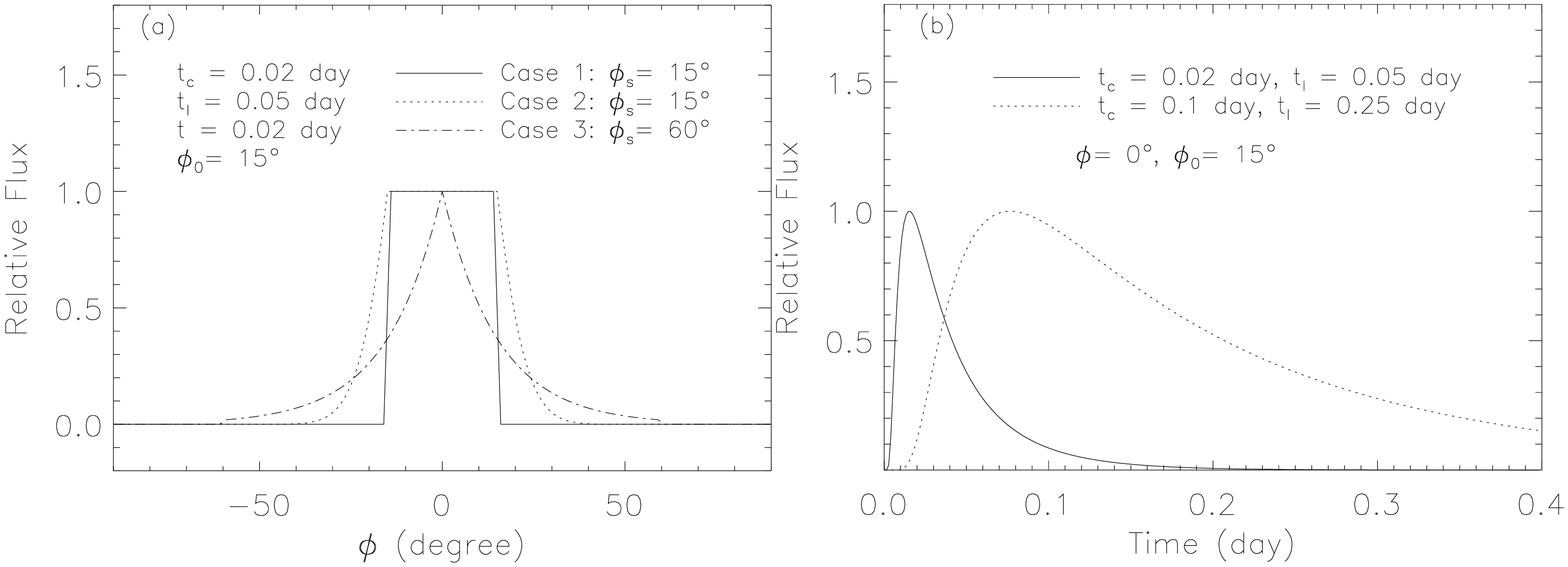} \caption{Comparison of
proton fluxes with different source models and source durations.
\label{source}}
\end{figure}

Here the particles are injected from the SEP source near the Sun. $\phi(\theta,\varphi)$ is the angle between the source center and
any point $(\theta,\varphi)$  near the Sun where the particles are injected.  We use three models to describe different scenarios: {\bf{ In case 1}} SEPs are accelerated by a flare, and particles do not propagate in the solar atmosphere; { \bf{In case 2}} SEPs are accelerated by a flare, and particles can propagate in the solar atmosphere; {\bf{And in case 3}} SEPs are accelerated by a coronal shock. The flare source model in case 1 and
2 is obtained {\bf{by}} following \cite{reid1964diffusive}, and the shock model in case 3 is obtained following \cite{Kallenrode1997JGR...10222311K}.
$H(x)$ is the Heaviside step function.
${\phi_s}$ is used to control the angular width of the source.
${\phi_0}$ describes how the source intensity decreases towards the flank of the source, {\bf{ and it is set to $15^\circ$ in the following simulations unless otherwise stated in the text.}}  $E_k$ is energy of the particles, and $\gamma$ is the spectral index of source particles. $t_c$ and $t_l$ are time constants to indicate the rise and decay
timescales, respectively. Here, we set a typical value of  $\gamma=3 $ for the spectral index of source particles.  Figure \ref{source} (a) shows spatial distributions of flux 
{\bf{at $t=0.02$ day}} normalized by the peaks in the cases of different source models (Cases 1, 2, and 3). Here, the parameters of source duration are set to $t_c=0.02$ day and $t_l=0.05$ day. $\phi_s$ is set to different {\bf{values}} corresponding to different source models. Figure \ref{source} (b) shows time profiles of flux {\bf{at $\phi=0^\circ$}}  normalized by the peaks in the cases of different source duration.  In our simulations,  the source rotates with the Sun rotation. 


We use a time-backward Markov stochastic process method to solve the
transport equation (\ref{dfdt}) \citep{Zhang1999ApJ...513..409Z}.
The initial-boundary value problem of the SEP transport equation can
be reformulated to stochastic differential equations, so it can be
solved by a Monte-Carlo simulation of Markov stochastic process, and
the SEP distribution function can be derived. In this method, we
trace particles from the observation point back to the injection
time from the SEP source. Only those particles in the source region
at the initial time contribute to the statistics. For detailed
description of the method, please refer to
\citet{Qin2006JGRA..11108101Q}.

\section{RESULTS}

Note that the inner boundary is $0.05$ AU, and {\bf{the}} outer  boundary is $50$ AU.
We use two different parallel mean free paths ${\lambda _\parallel } = 0.126 $ AU  and ${\lambda_\parallel } = 0.3$ AU for $10$ MeV protons {\bf{ in the ecliptic}}  at $1$ AU.
Based on the simulation
results of \cite{diaz2011delay}, interplanetary scattering
has great effect on the derived solar release time of VDA method in the case of ${\lambda _\parallel } = 0.126$ AU, while
it has only a small effect on the derived solar release time
in the case of ${\lambda _\parallel } = 0.3$ AU. As a result,  
${\lambda _\parallel } = 0.126$ AU is used {{\bf as strong scattering }} in the interplanetary space, and   ${\lambda _\parallel } = 0.3$ AU is used as  weak scattering. In the following simulations, $\lambda_\parallel$ is set to $0.126$ AU for $10$ MeV particles {\bf{ in the ecliptic}}  at $1$ AU  {\bf{ in the following simulations unless otherwise stated in the text}}. We choose different $t_c$ and $t_l$ to study different duration of the source. We set $t_c = 0.02$ day ($0.48$ hour) and $t_l = 0.05$ day ($1.2$ hours) as {\bf{an {\bf{impulsive}} duration}}, and set
 $t_c = 0.1$ day ($2.4$ hours) and  $t_l = 0.25$ day ($6$ hours) as a {\bf{gradual}} duration case.

Before we can determine the detected onset time from the
simulated time profiles, we have to define a background level of the
flux. In a real SEP event, this background may be either due to the
level of galactic cosmic rays or previous SEP event. In this paper,
we choose the background level as a constant fraction $A$ of the maximum
intensity. In each energy channel, we set the background fraction $A$
as $10^{-5}$, $10^{-3}$, and $10^{-1}$, corresponding to a low background
level, middle background level, and high background level, respectively.

\subsection{Particles  Not Propagating in the Solar Atmosphere }
In this subsection,  we use the source model as shown in 
the case 1 of Equation (\ref{boundaryCondition}), i.e., particles
are accelerated by a flare without propagating in the  solar atmosphere. In addition, $\phi_s$ is set to $15^\circ$.

\subsubsection{Effect of Perpendicular Diffusion on the Onset Time of SEP Event}
\begin{figure}
\epsscale{1.} \plotone{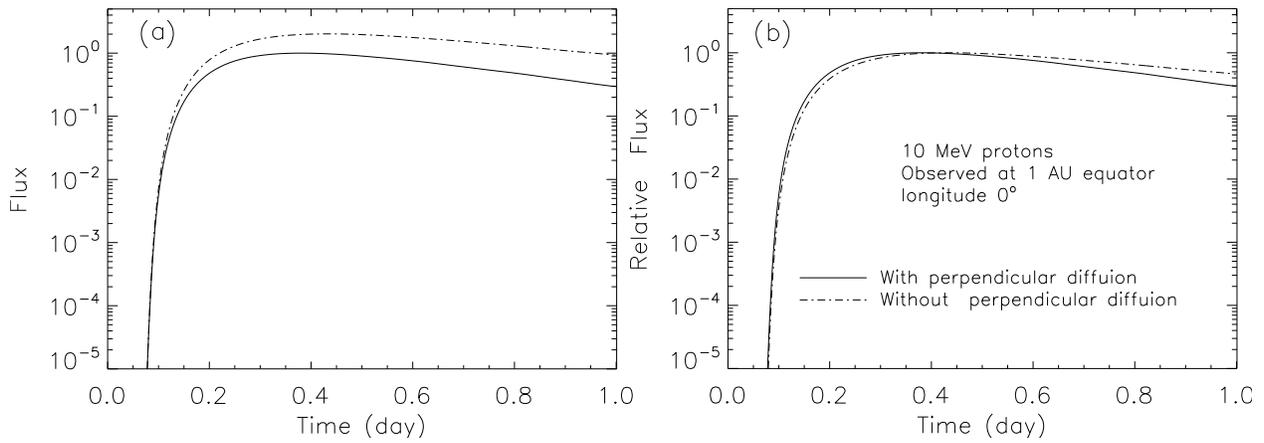} \caption{Comparison of
$10$ MeV proton fluxes with perpendicular diffusion (solid line) and
without perpendicular diffusion (dashed-dotted line).
\label{with_without_per}}
\end{figure}

Figure \ref{with_without_per} (a) shows time profiles of $10$ MeV
protons' omnidirectional flux, in the cases with and without
perpendicular diffusion. The source duration is set as a {\bf{gradual}} one, i.e., $t_c = 0.1$ day and
$t_l = 0.25$ day as a {\bf{gradual}} duration. The solid line indicates the case with
perpendicular diffusion, and the dashed-dotted line indicates the case without perpendicular diffusion. The observer is located at  {\bf{ in the ecliptic}} {\bf{ at 1 AU}} and
$0^\circ$ longitude, and the observer's field line is connected directly to the center of source near the Sun.
Comparing the time
profiles of flux, we can find {\bf{that}} the observed flux is smaller with the
perpendicular diffusion. {\bf{The reason is that,}} in this situation, the particles can leave field lines {\bf{because of perpendicular diffusion}}. Figure
\ref{with_without_per} (b) shows the time profiles of flux normalized by the peaks.   
As one can see, the
onset times are much the same with and without perpendicular diffusion. {\bf{So}} the two cases could not  be distinguished observationally.
We also show the normalized  flux  {\bf { to study the onset time}} in the following cases.

\subsubsection{Observers at Different Locations}
\begin{figure}
\epsscale{1.} \plotone{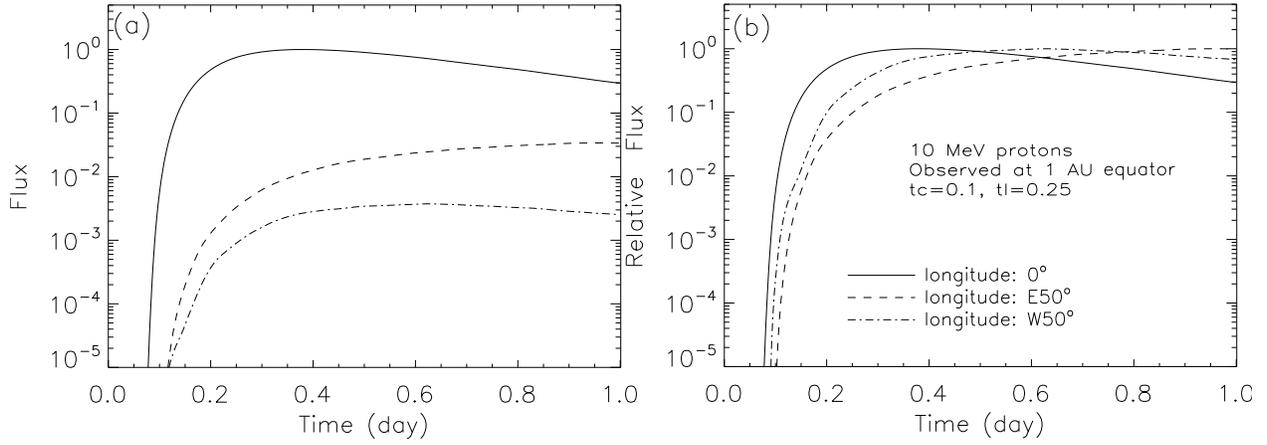} \caption{Comparison of $10$
MeV proton fluxes observed at different locations.  
\label{CM_E60_E50}}
\end{figure}

Figure \ref{CM_E60_E50} (a) shows time profiles of $10$ MeV protons' omnidirectional  flux detected by three observers.
The source duration is set as a {\bf{gradual}} one. 
The observers are located {\bf{ in the ecliptic}} with different longitudes, so that the center of source is located at $0^\circ$, $50^\circ$ west, and $50^\circ$ east  to the {\bf{foot-point}} of the observer, which are labeled as $0^\circ$ (solid line), $W50^\circ$ (dashed-dotted line), and $E50^\circ$ (dashed line), {\bf{respectively}}. When an observer is connected to the center of source directly by {\bf{the}} IMF, energetic particles can arrive at the observer’s location by following the field lines.
In the cases of $W50^\circ$ and $E50^\circ$, the two observers' field lines are disconnected from the SEP source, because the half width of source is only $15^\circ$. Therefore  energetic particles can only be detected by the observers with the effect of perpendicular diffusion during the onset time.  According to the {\bf{above}} three observers, the peak of flux is the largest when the observer is connected directly to the source by {\bf{the}} IMF, and it is the smallest when the center of source is located at $50^\circ$ west to the IMF {\bf{foot-point}} of the observer.
Due to the effect of convection, the particles rotate with the Sun after they are emitted. 
More particles are injected to the field line of observer if the source is located at $50^\circ$ east than that at $50^\circ$ west.
As a result, the flux of $W50^\circ$ is smaller than that of $E50^\circ$. This effect would lead to {\bf{the}} east-west asymmetry of SEPs distribution in the interplanetary space. Figure \ref{CM_E60_E50} (b) shows time profiles of normalized omnidirectional flux for the cases in Figure \ref{CM_E60_E50} (a). According to the observers, the onset time is {\bf{the}} earliest when the observer is connected to the source, and it is {\bf{the}} latest  when the center of source is located at $50^\circ$ east to the IMF foot-point of the observer.


\subsection{Different Source and Transport Models}

\begin{figure}
\epsscale{0.7} \plotone{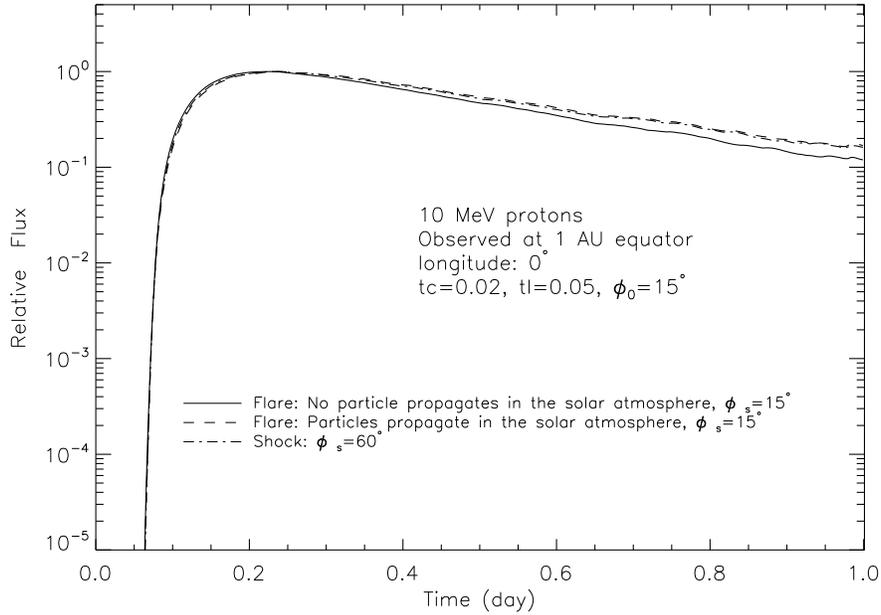} \caption{Comparison of $10$ MeV proton fluxes observed at $1$ AU which are produced by different SEP sources. The solid line indicates the case when particles are accelerated 
by {\bf{a}} flare, and particles do not propagate in the solar atmosphere. 
The dashed line indicates the  case when  particles are accelerated by a flare, and particles can propagate in the solar atmosphere.
The dashed-dotted line indicates the  case when particles are accelerated by a coronal shock.
\label{compare_diffuse_and_no_diffuse}}
\end{figure}

Figure \ref{compare_diffuse_and_no_diffuse} shows time profiles of   $10$ MeV protons' normalized omnidirectional flux in the cases with different source models.
The solid line  indicates that particles do not propagate in the solar atmosphere with case 1 of Equation (\ref{boundaryCondition}) ($\phi_s=15^\circ$). 
The dashed line  indicates that particles can propagate in the solar atmosphere with case 2 of Equation (\ref{boundaryCondition}) ($\phi_s=15^\circ$).
The dashed-dotted line  indicates that particles are accelerated by a coronal shock with case 3 of Equation (\ref{boundaryCondition}) ($\phi_s=60^\circ$).
In all of the three cases, the observer is located {\bf{ in the ecliptic}}  at $1$ AU, and is connected to the center of
source. The time profiles are similar during
the rising phase in all cases. As a result, when the observer is connected to the center of source, the spatial distribution of source does not affect the VDA results.

\begin{figure}
\epsscale{1.} \plotone{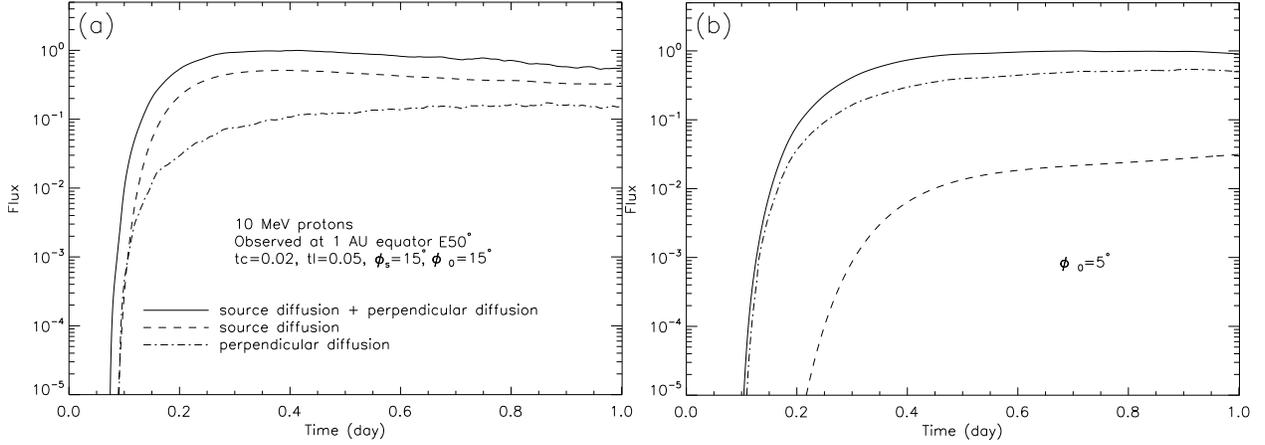} \caption{Comparison of $10$ MeV proton fluxes in the cases with different propagation models.  The solid lines indicate
the case when particles can propagate in the solar atmosphere, and can also cross the field lines in the interplanetary space with perpendicular diffusion. The dashed lines indicate the case when particles can propagate in the solar atmosphere, but without perpendicular diffusion in the interplanetary space. The dashed-dotted lines indicate the  case when  particles can  cross the field lines in the interplanetary space with perpendicular diffusion, but without propagation in the solar atmosphere.
\label{E50_with_different_source_diffusion}}
\end{figure}

Figure \ref{E50_with_different_source_diffusion} shows time profiles of  $10$ MeV protons' omnidirectional flux in the cases with different propagation models. 
In all  cases, the observer is located {\bf{ in the ecliptic}} at $1$ AU, and {\bf{ the center of 
SEP source is E$50^\circ$ to the observer  with $\phi_s = 15^\circ$.}}  The source parameter $\phi_0$ is set to  $15^\circ$ and $5^\circ$ in Figure \ref{E50_with_different_source_diffusion} (a) and (b), respectively.
The source model of solid lines and dashed lines is in case 2 of Equation (\ref{boundaryCondition}). 
The source model of dashed-dotted line is in case 1 of Equation (\ref{boundaryCondition}), which indicates that particles do not propagate in the solar atmosphere.
In addition, the solid and dashed-dotted lines indicate that particles propagate in the interplanetary space with perpendicular diffusion, while the dashed lines indicate  particles  without perpendicular diffusion. 
In Figure \ref {E50_with_different_source_diffusion} (a), 
the flux indicated by the solid line is the largest,
and that indicated by the dashed-dotted line is the smallest.
In Figure \ref {E50_with_different_source_diffusion} (b), 
the flux indicated by the solid line is the largest,
and that indicated by the dashed line is the smallest.
{\bf{Therefore, the propagation effect of particles in 
the solar atmosphere is stronger/weaker than that of
the perpendicular diffusion in the interplanetary space,
when the particle source  decreases slower/faster towards the flank of the source with $\phi_0=15^\circ$/$\phi_0=5^\circ$.}}
Comparing these two panels, the time scale of rising phase of flux indicated by solid line in the panel (a) is  smaller than that in the panel (b).




\subsection{Particles Propagating in the Solar Atmosphere}

\begin{figure}
\epsscale{1.} \plotone{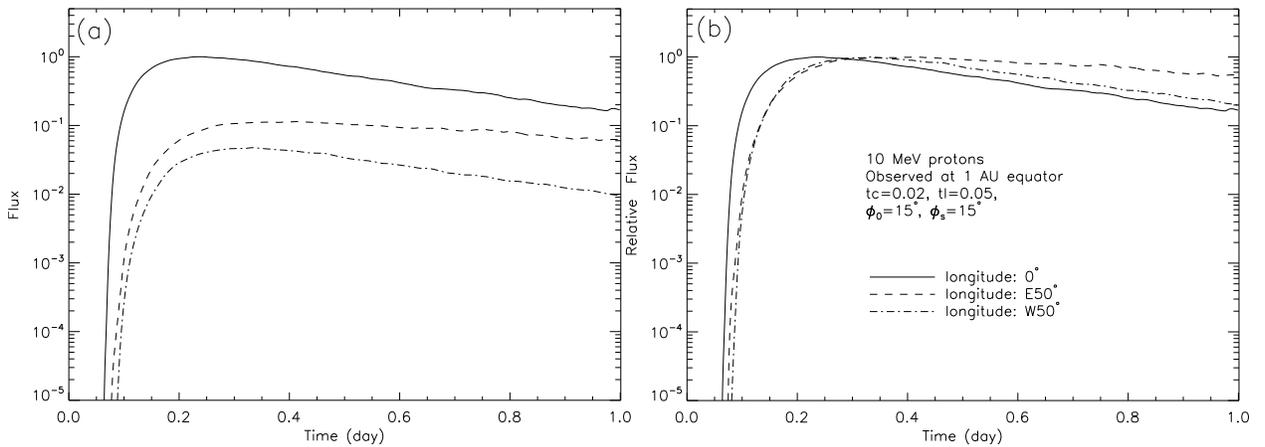} \caption{Comparison of $10$ MeV proton fluxes observed at different locations.  The particles are accelerated by a flare, and  the $\phi_s$ is set to  $15^\circ$.
\label{flare_CM_E50_W50}}
\end{figure}

Figure \ref{flare_CM_E50_W50} (a) shows time profiles of $10$ MeV protons' omnidirectional flux observed at different locations, while Figure \ref{flare_CM_E50_W50} (b) shows the normalized flues. 
We use the source model as
shown in the case 2 of  Equation (\ref{boundaryCondition}), and $\lambda_\parallel$ is set to $0.126$ AU for $10$ MeV particles at $1$ AU equatorial plane.
Here, particles produced by flare can propagate in the solar atmosphere, and they can also cross field lines
with perpendicular diffusion in the interplanetary space.
The source parameter $\phi_s$ is set to $15^\circ$.
Three observers are located {\bf{ in the ecliptic}} at $1$ AU  with different longitudes, so that the center of source is located at {\bf{ $0^\circ$, $W50^\circ$, and $E50^\circ$, respectively, to the foot-point of the observer.}}
According to the  observers, the flux is the largest when 
the observer is connected to the center of source, and is 
the smallest when the center of source is located at $W50^\circ$ 
to the {\bf{foot-point}} of the observer.
Figure \ref{flare_CM_E50_W50} (b) reveals that the onset times of three fluxes are different. 
In the source model, when ${\phi}$ is larger than $15^\circ$,  
the time profile of SEP source changes with  ${\phi}$.
As a result, the time profiles of SEP flux in cases of $E50^\circ$ and $W50^\circ$ increase more slowly than that in the case of longitude $0^\circ$. {\bf{In this condition}}, 
when the observer is far from the center of source ($\phi>15^\circ$), the angular distance should affect the VDA results.


\subsection{Particles Accelerated by a Large Corona Shock}
\begin{figure}
\epsscale{1.} \plotone{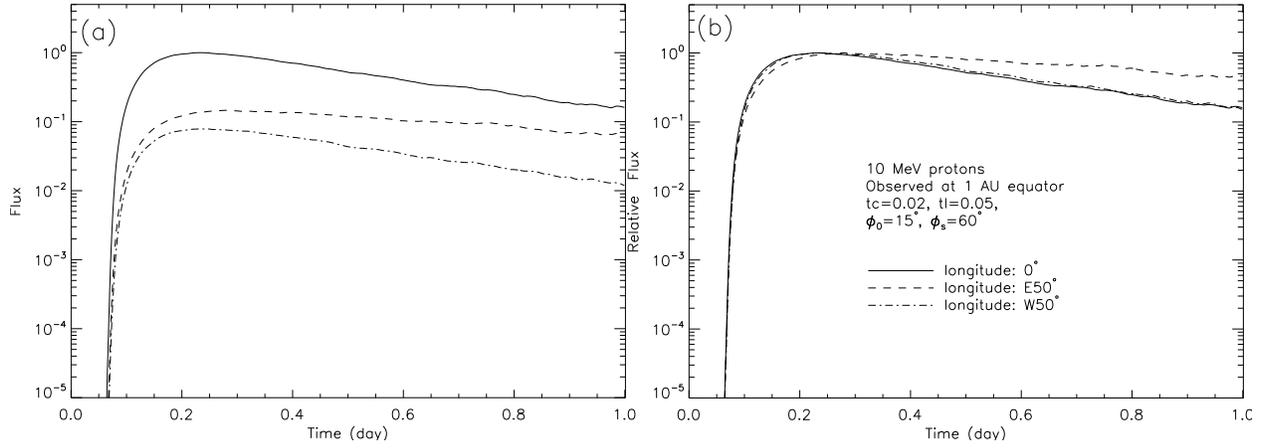} \caption{Comparison of $10$ MeV proton fluxes observed at different locations.  The particles are accelerated by a coronal shock, and  the $\phi_s$ is set to  $60^\circ$.
\label{shock_CM_E50_W50}}
\end{figure}

Figure \ref{shock_CM_E50_W50} is similar to Figure \ref{flare_CM_E50_W50} but with different SEP source near the Sun. 
We use the SEP source model as
shown in the case 3 of Equation (\ref{boundaryCondition}), and $\lambda_\parallel$ is set to $0.126$ AU for $10$ MeV particles at $1$ AU equatorial plane. Particles are accelerated by a corona shock, and they can also cross the field lines with perpendicular diffusion in the interplanetary space. The source parameter $\phi_s$ is set to $60^\circ$, and $\phi_0$ is set to $15^\circ$. 
Three observers are located at $1$ AU equatorial plane with different longitudes, so that the center of source is located at {\bf{ $0^\circ$, $W50^\circ$, and $E50^\circ$, respectively, to the foot-point of the observer.}}
In this case, the observers are connected to the  source by magnetic field lines. Therefore, the effects of particles propagating in the solar atmosphere and perpendicular diffusion in the interplanetary space have little {\bf{effect}} on the onset time of SEP fluxes.
Since the time profile of source does not change with $\phi$ in our model, the time profiles of fluxes detected by three observers are similar during the rising phase in Figure \ref{shock_CM_E50_W50} (b).  
As a result, the angular distance between the center of source and  {\bf{the}} observer should not affect the VDA results. 
However, if the time profile of SEP source changes with angular distance, the time profiles of flux detected by the observers should be different. 
This conclusion could be deduced from the results in  Figure \ref{flare_CM_E50_W50}. In this case, the VDA results should change
with the angular distance between the center of source and observer.

\subsection{VDA Method Results}

In this subsection, we will study how {\bf{the}} perpendicular diffusion and different source models affect the VDA method results. The VDA method assumes that the first observed particles are the {\bf{ones}} travelling  along the magnetic field lines, and the path length {\bf{travelled}} by particles is independent on energy. If this holds, the transport time for SEPs is given by

\begin{equation}
{t_o} - {t_i} = \frac{L}{v} \label{VDA},
\end{equation}

here, $L$ is a constant which represents the field line length. $t_o$ is the onset time of SEP flux which depends on particle speed. $t_i$ is a constant which represents release time of particles on the source. $v$ is {\bf{the}} speed of energetic particles.


Figure \ref{VDA} (a) and (b) show the dispersion of the onset time
changes with c/v according to different source durations, where the observers are located at equator $0^\circ$ longitude, and the observer's field line is connected directly to
the center of source near the Sun. The only difference between
Figure \ref{VDA}  (a) and (b) is the source duration times.
 In Figure \ref{VDA}, source model is set as the case 1 of Equation \ref{boundaryCondition}, and the source parameter $\phi_s$ is set to  $15^\circ$.
Here, we get time profiles of SEP fluxes with simulations for four energy channels, $10$ MeV, $20$ MeV, $40$ MeV, and $80$ MeV.
We set the SEP background as $10^{-5}$ of the flux peak {\bf{indicating a low background}}.
{\bf{From SEP fluxes}}, we obtain the
onset times as the times when fluxes rise above the background. As one can see, the onset time increases
linearly with c/v. Based on the results of data fitting, the release time near the Sun and {\bf{the}} interplanetary field length can be derived.  


With different observing locations, background levels, and source duration times, the calculated source release times and path lengths from the VDA method with simulation data are listed in Table \ref{strongScattering}.  In our simulations, the source release time is set to $0$, and the IMF is set to Parker field model with the solar wind speed $400$ km/s. In this table, the source release time and path length derived from VDA are labeled as $t_i$ and $L$, receptively.  ${\lambda_\parallel }$ is set to $ 0.126$ AU for $10$ MeV protons  {\bf{in the ecliptic}} at $1$ AU. Source model is 
set as the case 1 of Equation \ref{boundaryCondition}, and $\phi_s$ is set to  $15^\circ$.
The observers are all located at $1$ AU equatorial plane, but at different longitudes.
 When the observer is connected  to the source, we have the following findings. In the cases of {\bf{impulsive}} source
duration (cases 1, 2, and 3), the VDA release times are very close to the injection
times on the source (less than 3 minutes).
In the cases of {\bf{gradual}} source duration (cases 4, 5, and 6), however, the VDA release times are
much later than the  real release times.
On the other hand, when the observer is disconnected  from the source, the onset time of SEP flux is later than the case  when the  observer is connected  to the source. Therefore, the VDA
results are generally much worse when the observer is disconnected  from the source with some exceptions.
For example, in
$W50^\circ$  of case 4, the release time of VDA result is very close to the injection time on
 the source (less than $2$ minutes). Obviously, this result is obtained
fortuitously, and it can not be taken as an indication that the VDA method is valid.
In all six cases ({\bf{cases}} 1, 2, 3, 4, 5 and 6), the path lengths obtained from VDA
method are longer than the real ones from Parker spiral.
Even when the observer is connected to the center of flare with {\bf{impulsive}} duration source and low background SEP level, the path length is still larger than that of Parker spiral. This is because  scattering in the interplanetary space could not
be ignored due to the value of mean free path  used in our simulation. {\bf{Besides}}  scattering in the interplanetary space, the source duration, source location, and SEP background level also affect the path length result of VDA. Therefore, in the case 6 of Table \ref{strongScattering}, the path lengths are much larger than that of Parker spiral.


Table \ref{weakScattering} is similar to Table \ref{strongScattering} except {\bf{the}} mean free path.  ${\lambda_\parallel }$ is set to $ 0.3$ AU for $10$ MeV protons  {\bf{in the ecliptic}} at $1$ AU  in Table \ref{weakScattering}.  When the observer is connected  to the source, we have the following findings. In the cases  1 and 2, the VDA release times are very close to the injection times on the source (less than 3 minutes). However,  in case 3 the difference between VDA release time and the injection times can be as large as $6.8$ minutes. 
In the cases of {\bf{gradual}} source duration (cases 4, 5, and 6),  the VDA release times are much later than the  real release times.
Due to  the larger mean free paths, the path length derived from VDA in case 1 is much smaller than that in Table \ref{strongScattering}. The value $1.34$ AU is {\bf{ closer}} 
to the length of Parker spiral.
On the other hand, when the observer is disconnected  from the source, the VDA release times are also very close to
the injection times (less than 3 minutes) in case 1. 
However, in other cases, the VDA release times are generally worse when the observer is disconnected  from the source with some exceptions. In all six cases ({\bf{cases}} 1, 2, 3, 4, 5 and 6), the path lengths obtained from VDA method are longer than the real ones from Parker spiral. Comparing Table \ref{weakScattering}  with Table \ref{strongScattering}, the  path lengths obtained from VDA method in 
Table \ref{weakScattering}  are smaller than that in Table \ref{strongScattering}.


Table \ref{sourceDiffusion} is similar to Table \ref{strongScattering} except the source model and mean free path.
In Table \ref{sourceDiffusion}, the source model is set as the case 2 of Equation \ref{boundaryCondition},
and particles can diffuse at the source region.
The source parameters are set to $\phi_s=15^\circ$, ${t_c=0.02}$ day, and ${t_l=0.05}$ day in all six cases. When  ${\lambda_\parallel }$ is set to $ 0.126$ AU  for $10$ MeV protons  {\bf{in the ecliptic}} at $1$ AU, the {\bf{propagation}} effect of particles  in the solar atmosphere is stronger than that of {\bf{the}} perpendicular diffusion in the interplanetary space. 
In cases 1, 2, and 3,  
${\lambda_\parallel }$ is set to $ 0.126$ AU for $10$ MeV protons, while ${\lambda_\parallel }$ is set to $ 0.3$ AU for $10$ MeV protons in cases 4, 5, and 6. 
Comparing the cases 1, 2, and 3 in Table \ref{sourceDiffusion}  with the {\bf{impulsive}} duration cases in Table \ref{strongScattering}, we have the following findings. The VDA release times are much closer to the injection times on the source in Table \ref{sourceDiffusion} than that in Table \ref{strongScattering}, and the  path lengths in Table \ref{sourceDiffusion}  are generally smaller than that in Table \ref{strongScattering}. The difference between the VDA release times and the injection times are within $7$ minutes. Comparing the cases 4, 5, and 6 in Table \ref{sourceDiffusion}  with the {\bf{impulsive}} duration cases in Table \ref{weakScattering}, we find that the VDA results are similar in these two tables especially in the low background case. This is because  when  ${\lambda_\parallel }$ is set to $ 0.3$ AU for $10$ MeV protons at 1 AU equatorial plane, the effect of perpendicular diffusion in the interplanetary space is stronger than  that of particles propagation in the solar atmosphere.

\section{DISCUSSION AND CONCLUSIONS}
In this paper, we discuss the uncertainty of the simple assumptions of VDA method.
Firstly, the VDA results could be significantly {\bf{affected}} by interplanetary scattering. \citep{Kallenrode1990ICRC5229K,Lintunen2004A&A420343L,diaz2011delay}.
Secondly, the onset time of SEP event is hard to be determined in practical
applications, since it can
be significantly delayed by the background level
\citep{Saiz2005ApJ...626.1131S}. Thirdly, particles can cross the
field lines when they transport in the space. The perpendicular
diffusion plays a very important role in the release time determination, especially when the observer's field line is disconnected from the source
\citep{Zhang2009ApJ...692..109Z,Qin2011ApJ73828Q,he2011propagation}.
Fourthly, different source models affect the accuracy of results of VDA method. For example, particles accelerated by a flare may directly propagate in the solar atmosphere \citep{wibberenz1989coronal}; and a large  shock could provide a very wide source \citep{mason1984temporal,Gosling1993The,Zank2000JGR...10525079Z,li2003energetic}.

By numerically solving the focused Fokker-Planck equation, we
have calculated SEPs' intensity time profiles including the
perpendicular diffusion. We set
different source duration and background level to study the onset
times observed by observers. Comparing the time profiles of SEP
fluxes observed at different locations, we have studied the effect
of different source models and perpendicular diffusion on the onset times of SEP events, and its
influence on the VDA method results.  Our new findings are the following.

(1) If SEPs are produced by a solar flare, they can spread much wider than the source region by two possible mechanisms. 
In the first one, particles  propagate  in the solar atmosphere. In the second one, particles cross the field lines in the interplanetary space with perpendicular diffusion. 
In this case, the VDA results can be affected by the above two mechanisms when the observer is not connected to the source at {\bf{the}} initial time. In addition, due to the effect of convection, more particles are injected in the field line of observer when {\bf{the}}  source is located at {\bf{the}} east flank to the {\bf{foot-point}} of observer than that in the case of {\bf{the}} west flank. This effect would lead to {\bf{the}} east-west asymmetry of SEPs distribution in the interplanetary space.

(2) When the observer is connected to the source by {\bf{the}} IMF, comparing the time profiles of fluxes in the cases with and without  perpendicular diffusion, or with and without particles  propagation in the solar atmosphere, the onset times are much the same. 
In  this case, {\bf{ the effects of particles}} perpendicular diffusion in the interplanetary space  and propagation in the solar atmosphere do not affect the results of VDA significantly.  
The results obtained by previous simulations, which  didn't include these two mechanisms  \citep{Kallenrode1990ICRC5229K,Lintunen2004A&A420343L,Saiz2005ApJ...626.1131S,
diaz2011delay}, still holds when these two mechanisms are included.


(3) If SEPs accelerated by a solar flare can not 
propagate in the solar atmosphere, the SEP source region 
is about the size of solar flare.
In this case, when the observer is disconnected from the source by the IMF, the energetic particles can be detected with the effect of the perpendicular diffusion. The onset
time is later when the observer is disconnected from the source than that when the observer is connected to the source.
In the cases of weak scattering, the solar release time derived
from VDA method is close to the injection time when the observer is disconnected from the source with impulsive source duration
and low background.
However, in the cases of strong scattering, the release time and the path length obtained from the VDA method are much different  from the real values except some fortuitous cases.


(4)  If SEPs accelerated by a solar flare can  
propagate in the solar atmosphere, the SEP source region 
should be larger than the size of solar flare as times go by. 
When the observer is far from the SEP source at the initial time, particles will spend some time on leaving the source to the observer's field line. As a result, the time scale of rising phase of flux is larger than that in the case when the observer is connected to the center of source. When source diffusion in the solar atmosphere is faster than perpendicular diffusion in the interplanetary space, the solar release time derived from VDA method is close to injection time when the observer is disconnected from the source with low background. If source diffusion in the solar atmosphere is slower than perpendicular diffusion, the VDA  results are significantly affected by the mean free path. In the case of weak scattering, the solar release time derived from VDA method is still valid with impulsive source duration and low background. However, in the case of strong
scattering, the solar release time derived from VDA method is
not valid.

(5) If SEPs are accelerated by a large scale corona shock,
the source can cover {\bf{a}} very wide region due to the
size of corona shock. The observers
located at different locations can be connected to the Sun simultaneously. Therefore, the effects of particles propagating in the solar atmosphere and perpendicular diffusion in the interplanetary space have little {\bf{effect}} on the onset time of SEP fluxes and the VDA results, and the accuracy of VDA {\bf{method}} depends on other conditions.
If the time profile of SEP source does not change with the angular distance between the foot-point of observer's magnetic field line and
the center of the source, the onset time of fluxes observed at different locations could be much the same.  In this case, the VDA results do not change with the angular distance.
Otherwise,  the time profile of SEP source changes with the angular distance,  the onset time of SEP flux and the VDA results will  change with {\bf{the angular distance.}}

(6) In our simulations,  the VDA results could be  significantly {\bf{affected}}  by the location and  size  of SEP source. As it is shown in previous studies \citep{Kallenrode1990ICRC5229K,Lintunen2004A&A420343L,Saiz2005ApJ...626.1131S,
diaz2011delay}, the VDA results are also significantly affected by the time profile of source, parallel mean free path, and background level. 
In order to reduce error in the results of VDA method, an ideal SEP
event should meet the following conditions, such as, {\bf{impulsive}} source duration, large parallel mean free path, low background level, and good connection between the observer and the source.

\acknowledgments
The authors thank the anonymous referee for valuable comments. We are partly supported by grants   NNSFC 41374177,  NNSFC
41125016, and NNSFC 41304135, the CMA grant GYHY201106011, and the Specialized Research Fund for State
Key Laboratories of China.
The computations were performed by Numerical Forecast Modeling R\&D
and VR System of State Key Laboratory of Space Weather and Special
HPC work stand of Chinese Meridian Project.


\begin{figure}
\epsscale{1.} \plotone{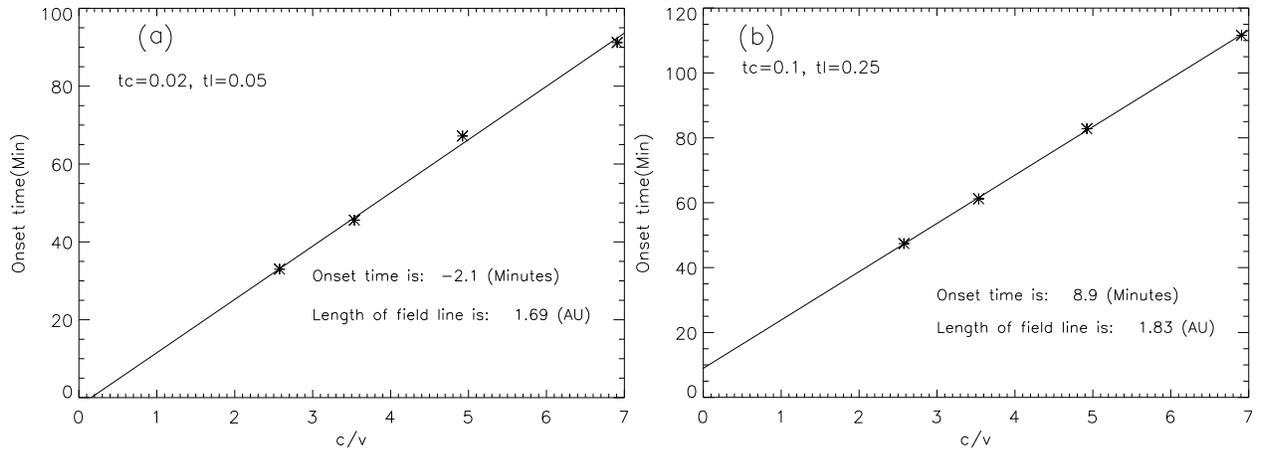} \caption{The results of  the SEP release time and path length {\bf{derived from the VDA method}}.
\label{VDA}}
\end{figure}

\clearpage

\begin{table}
\caption {Results of VDA method with strong scattering*. \label{strongScattering}}
\begin{tabular} {|l|l|l|l|l|l|} \tableline
Case & Duration & Background & Location & $t_i$ (min) & L (AU)
\\ \tableline\tableline

\multirow{3}{*}{1}&  &   & Center & -2.1 &  1.69 \\
&{\bf{Impulsive}} & Low &   $W50^\circ$ & -12 &  2.42  \\
&  &  &  $E50^\circ$ & -18 &  2.51 \\
\tableline\tableline

\multirow{3}{*}{2}&  &   & Center & -1.7 &  1.86 \\
&{\bf{Impulsive}} & Middle &  $W50^\circ$ & -8.3 &  2.38\\
& &  &  $E50^\circ$ & -18 &  2.68\\
\tableline\tableline

\multirow{3}{*}{3}&  &   & Center & -0.98 &  2.34 \\
&{\bf{Impulsive}} & High &  $W50^\circ$ & -23 &  3.95\\
& &  &  $E50^\circ$ & -43 & 4.83\\
\tableline\tableline

\multirow{3}{*}{4}&  &   & Center &  8.9 &  1.83\\
&{\bf{Gradual}} & Low & $W50^\circ$ &  1.5 &  2.39 \\
& &  & $E50^\circ$ & -8.9 &  2.70\\
\tableline\tableline

\multirow{3}{*}{5}&  &   & Center &  12 &  2.09 \\
&{\bf{Gradual}} & Middle &  $W50^\circ$ & 9.9 &  2.66 \\
& &  & $E50^\circ$ & -7.7 &  3.34 \\
\tableline\tableline

\multirow{3}{*}{6}&  &   & Center &  26 & 2.98 \\
&{\bf{Gradual}} & High &  $W50^\circ$ & 26 &  4.65 \\
& &  & $E50^\circ$ & -6.9 &  6.45 \\
\tableline

\end{tabular}

\tablenotetext{*}{In this table, ${\lambda_\parallel }$ is set to $ 0.126$ AU for $10$ MeV protons {\bf{in the ecliptic}} at $1$ AU.}
\end{table}

\begin{table}
\caption {Results of VDA method with weak scattering*. \label{weakScattering}}
\begin{tabular} {|l|l|l|l|l|l|} \tableline
Case & Duration & Background & Location & $t_i$ (min) & L (AU)
\\ \tableline\tableline

\multirow{3}{*}{1}&  &   & Center & 2.8 &  1.34 \\
&{\bf{Impulsive}} & Low &  $W50^\circ$ & -1.3 &  1.60  \\
&  &  &  $E50^\circ$ & 2.1 &  1.39 \\
\tableline\tableline

\multirow{3}{*}{2}&  &   & Center & 2.6 &  1.46 \\
&{\bf{Impulsive}} & Middle &  $W50^\circ$ & 6.1 &  1.52\\
& &  &  $E50^\circ$ & 4.0 &  1.41\\
\tableline\tableline

\multirow{3}{*}{3}&  &   & Center & 6.8 &  1.70 \\
&{\bf{Impulsive}} & High &   $W50^\circ$ & 6.9 & 1.87 \\
& &  &  $E50^\circ$ & 6.0 & 1.84\\
\tableline\tableline

\multirow{3}{*}{4}&  &   & Center &  11 &  1.48\\
&{\bf{Gradual}} & Low &  $W50^\circ$ &  13 &  1.57 \\
& &  & $E50^\circ$ & 14 &  1.43\\
\tableline\tableline

\multirow{3}{*}{5}&  &   & Center &  16 &  1.61 \\
&{\bf{Gradual}} & Middle &  $W50^\circ$ & 17 &  1.78 \\
& &  &  $E50^\circ$ & 20 &  1.63 \\
\tableline\tableline

\multirow{3}{*}{6}&  &   & Center &  38 & 1.90 \\
&{\bf{Gradual}} & High &  $W50^\circ$ & 44 &  2.16 \\
& &  & $E50^\circ$ & 42 &  2.67 \\
\tableline

\end{tabular}
\tablenotetext{*}{In this table, ${\lambda_\parallel }$ is set to $ 0.3$ AU for $10$ MeV protons {\bf{in the ecliptic}} at $1$ AU.}
\end{table}

\begin{table}
\caption {The effects of source diffusion in the VDA method.\label{sourceDiffusion}}
\begin{tabular} {|l|l|l|l|l|l|} \tableline
Case & Scattering & Background & Location & $t_i$ (min) & L (AU)
\\ \tableline\tableline

\multirow{3}{*}{1}&  &   & Center & -2.1 &  1.69 \\
& Strong$^a$ & Low & $W50^\circ$ & -0.91 &  1.91   \\
&  &  & $E50^\circ$ &  -4.8 &  1.96 \\
\tableline\tableline

\multirow{3}{*}{2}&  &   & Center & 0.41 &  1.82 \\
& Strong$^a$ & Middle & $W50^\circ$ &  -6.9 &  2.41 \\
& &  &  $E50^\circ$ &  -5.9 &  2.31\\
\tableline\tableline

\multirow{3}{*}{3}&  &   & Center & -0.88 &  2.38 \\
& Strong$^a$ & High &  $W50^\circ$ & 2.8 &  3.29\\
& &  &  $E50^\circ$ &  -0.72 & 3.34\\
\tableline\tableline

\multirow{3}{*}{4}&  &   & Center & 2.8 &  1.34 \\
& Weak$^b$ & Low & $W50^\circ$ & 0.4 &  1.56   \\
&  &  &  $E50^\circ$ &  2.1 &  1.39 \\
\tableline\tableline

\multirow{3}{*}{5}&  &   & Center & 3.5 &  1.45 \\
& Weak$^b$ & Middle & $W50^\circ$ &  4.6 &  1.63 \\
& &  & $E50^\circ$ &  3.9 &  1.51\\
\tableline\tableline

\multirow{3}{*}{6}&  &   & Center & 8.3 &  1.69 \\
& Weak$^b$ & High &  $W50^\circ$ & 11 &  2.02\\
& &  & $E50^\circ$ & 11 & 1.91\\
\tableline
\end{tabular}
\tablenotetext{a}{${\lambda_\parallel }$ is set to  $ 0.126$ AU for $10$ MeV protons {\bf{in the ecliptic}} at $1$ AU.}
\tablenotetext{b}{${\lambda_\parallel }$ is set to  $ 0.3$ AU for $10$ MeV protons {\bf{in the ecliptic}} at $1$ AU.}
\end{table}

\end{document}